\documentclass[conference]{IEEEtran}
\IEEEoverridecommandlockouts
\usepackage{cite}
\usepackage{amsmath,amssymb,amsfonts}
\usepackage{algorithmic}
\usepackage{graphicx}
\usepackage{textcomp}
\usepackage{xcolor}
\usepackage{url}
\usepackage{booktabs}
\usepackage[table]{xcolor}
\def\BibTeX{{\rm B\kern-.05em{\sc i\kern-.025em b}\kern-.08em
    T\kern-.1667em\lower.7ex\hbox{E}\kern-.125emX}}
\begin{document}

\title{Benchmarking Adult Addressee Classification Across Child- and Adult-Directed Speech Datasets
}

\if false
\author{
    \IEEEauthorblockN{1\textsuperscript{st} Given Name Surname}
    \IEEEauthorblockA{\textit{dept. name of organization (of Aff.)} \\
    \textit{name of organization (of Aff.)}\\
    City, Country \\
    email address or ORCID}
    \and
    \IEEEauthorblockN{2\textsuperscript{nd} Given Name Surname}
    \IEEEauthorblockA{\textit{dept. name of organization (of Aff.)} \\
    \textit{name of organization (of Aff.)}\\
    City, Country \\
    email address or ORCID}
    \and
    \IEEEauthorblockN{3\textsuperscript{rd} Given Name Surname}
    \IEEEauthorblockA{\textit{dept. name of organization (of Aff.)} \\
    \textit{name of organization (of Aff.)}\\
    City, Country \\
    email address or ORCID}
}
\fi
\author{
\IEEEauthorblockN{Sofoklis Kakouros}
\IEEEauthorblockA{\textit{Signal Processing Research Centre} \\
\textit{Tampere University}\\
Tampere, Finland \\
0000-0001-8996-0793}
\and
\IEEEauthorblockN{Daniil Kocharov}
\IEEEauthorblockA{\textit{Signal Processing Research Centre} \\
\textit{Tampere University}\\
Tampere, Finland \\
0000-0002-7858-5331}
\and
\IEEEauthorblockN{Okko R{\"a}s{\"a}nen}
\IEEEauthorblockA{\textit{Signal Processing Research Centre} \\
\textit{Tampere University}\\
Tampere, Finland \\
0000-0002-0537-0946}
}

\maketitle

\begin{abstract}
In this work, we present a comprehensive analysis of classification performance for distinguishing child-directed speech (CDS) from adult-directed speech (ADS) using speech data from corpora containing natural in-lab and in-the-wild CDS and ADS. We establish classification benchmarks for these datasets using self-supervised learning (SSL) representations, along with a range of time-pooled representations that go beyond first- and second-order statistics by incorporating cross-channel covariances in high-dimensional embeddings. In addition, we probe these representations to examine how different pooling methods capture prosodic information using linear probes. Overall, SSL-based representations prove particularly effective, achieving the best performance, while different pooling methods offer complementary advantages for the task.
\end{abstract}


\section{Introduction}
Child-centered daylong recordings (also called longform recordings), containing natural caregiver–child interactions, are of considerable interest across research areas such as developmental psychology, language acquisition, linguistics, and speech technology. At the same time, they pose significant challenges for analysis because they are collected in naturalistic settings, where acoustic conditions can vary greatly. Changes in environment, microphone position, speaker identity, overlapping speech, and background noise can all reduce audio quality and complicate interpretation \cite{li2026automated}.

Difficulties with naturalistic audio recordings become especially pronounced when the goal is to classify whether an adult is speaking to a child or to another adult (so-called addressee classification). In daylong recordings, this distinction is often difficult to determine because speech may be degraded, only partially captured, or embedded in noisy and highly dynamic interactions. As a result, the challenge lies not only in the variable quality of the recordings, but also in the complexity of the classification task itself, which relies on subtle acoustic, linguistic, and contextual cues.


Automatic child-directed speech (CDS)--adult directed speech (ADS) classification remains a relatively specialized problem, and the strongest published results are still mostly tied to specific corpora rather than showing broad transferability. The main ecologically valid benchmark for adult-addressee classification is the HomeBank Child/Adult Addressee Corpus (HB-CHAAC), introduced in the INTERSPEECH 2017 ComParE Addressee Challenge \cite{schuller2017interspeech}. HB-CHAAC contains conversations selected from several corpora \cite{bergelson2016seedlings, mcdivitt2016corpus, warlaumont2016corpus}. Across the literature, there is a clear gap between results obtained on higher-quality, controlled data \cite{schuster2014speaker} and those obtained on speech data from long-form recordings \cite{schuller2017interspeech, gosztolya17b_interspeech, akhtiamov2019cross,al2023hearttoheart}, highlighting the difficulty of the task under realistic conditions.

The best reported CDS--ADS results are still those from the ComParE Addressee Challenge baselines \cite{schuller2017interspeech}. In terms of unweighted average recall (UAR), the individual systems reached 60.1 for a CNN+LSTM, 67.7 for an SVM using eGeMAPS features, and 68.3 for an SVM using quantized eGeMAPS features. Combining these systems by selecting the prediction from the classifier with the highest summed confidence yielded the strongest reported performance at 70.2 UAR \cite{schuller2017interspeech}. Later approaches, including alternative RNN architectures \cite{gosztolya17b_interspeech}, data augmentation \cite{akhtiamov2019cross}, and adversarial training with multi-task learning \cite{al2023hearttoheart}, did not surpass this benchmark, with best results reaching 67.7 UAR. It is also important to note, that there is no standardized benchmark for this task.

A major reason for this limited progress is the nature of the data. CDS--ADS classification in realistic daylong recordings is especially challenging because the audio is often heavily degraded, for example in LENA-style recordings \cite{xu2008signal}, and includes variable recording conditions, multiple speakers, background noise, overlapped speech, sparse and ambiguous examples rendering labeling inaccurate \cite{cristia2020accuracy}, and class imbalance. These factors create substantial within-corpus variability and strong cross-corpus domain shifts \cite{cristia2021thorough,akhtiamov2019cross,lavechin2020open}. In addition, the acoustic differences between child- and adult-directed speech are not uniform: they can vary considerably across datasets, environments, and speakers, making it difficult for models to learn stable, transferable distinctions. Recent advances in self-supervised speech models \cite{baevski2020wav2vec, hsu2021hubert, chen2022wavlm} are especially relevant, as such models have outperformed earlier feature representations in many speech tasks \cite{amiriparian2024exhubert, keren2025omnilingual, charlot2025babyhubert}. This motivates evaluating self-supervised learning (SSL) approaches for CDS--ADS classification, particularly under the data-quality and domain-shift constraints that define the task.

In this work, we present baseline SSL and eGeMAPS results for adult addressee classification (ADS--CDS classification), using  both in-lab and long-form recordings. We evaluate both state-of-the-art SSL representations and conventional acoustic-prosodic features. We also examine a range of pooling strategies for capturing distinctions between these speech styles, with particular emphasis on covariance pooling—an application of image style transfer to speech \cite{gatys2016image}. Similar approaches \cite{kakouros2023speech,stafylakis2023extracting} have been shown to be effective for prosodic tasks such as emotion recognition \cite{kakouros25_interspeech}. Finally, we probe the learned representations and classifiers to better understand the underlying acoustic information they capture ---see also \cite{kakouros2026classification,kakouros2025sounding} for classifier-based identification of relevant eGeMAPS features. 

\section{Materials}

\subsection{ManyBabies dataset}
The ManyBabies dataset from the Infant-Directed Speech Preference study \cite{manybabies2020} was used to obtain natural CDS and ADS samples produced by the same speakers, as it includes in-laboratory recordings of 15 mothers speaking separately to their infants and to experimenters. Speech was collected using a lapel clip-on microphone connected to a smartphone. The conversations centered on five familiar objects (a ball, a shoe, a cup, a block, and a train) and five unfamiliar objects (a sieve, a globe, a whisk, a flag, and a bag of yeast). In total, the dataset comprises 1,281 CDS and ADS utterances, corresponding to 1.25~h of speech. The full dataset is used in this study.

\subsection{ACLEW dataset}
The ACLEW dataset---compiled by the \emph{Analyzing Child Language Experiences around the World} (ACLEW) project---consists of daylong, long-form and thereby naturalistic recordings from children aged 2--24 months with wearable audio recorders \cite{casillas2017babies}. The recordings were collected primarily using \textsc{LENA} devices \cite{xu2008signal}, with additional recording equipment used at some sites. In this work, we use the English subset of the dataset, consisting of four distinct corpora collected at different locations of US and UK \cite{bergelson2016seedlings,mcdivitt2016corpus,warlaumont2016corpus,vandam2016homebank}. From each recording, random 2-minute excerpts were sampled and then manually segmented, diarized by speaker type, and annotated as CDS or ADS, comprising a total of 9,909 utterances. Three of the corpora are available through HomeBank \cite{vandam2016homebank} and one through the Language Archive (\texttt{\url{https://archive.mpi.nl/}}). Additional project information is available at \texttt{\url{https://sites.google.com/view/aclewdid/home}}.

\section{Methods}

\subsection{Features}
We extract both handcrafted acoustic--prosodic features and self-supervised speech representations. For the former, we compute utterance-level descriptors using the eGeMAPS feature set implemented in openSMILE, yielding an 88-dimensional vector for each recording \cite{eyben2010opensmile,eyben2016egemaps}. To complement these low-dimensional summary descriptors, we additionally extract frame-level representations using three pretrained self-supervised speech encoders trained on English speech: \textbf{HuBERT-base} (\texttt{facebook/hubert-base-ls960}; 768-dimensional embedding) \cite{hsu2021hubert}, \textbf{WavLM-base-plus} (\texttt{microsoft/wavlm- base-plus}; 768-dimensional embedding) \cite{chen2022wavlm}, and \textbf{WavLM-large} (\texttt{microsoft/wavlm-large}; 1024-dimensional embedding) \cite{chen2022wavlm}. For eGeMAPS, a 20-ms frame length with a 10-ms hop size was used (the standard eGeMAPS configuration), while all SSL representations use a convolutional feature encoder with an effective frame length of 25-ms and a 20-ms hop size.

\textbf{WavLM-Base-Plus} is included in the experiments since its broader pretraining corpus and denoising-based objective could yield representations that are more robust to real-world variability and noise, making it a suitable choice for day-long recordings \cite{chen2022wavlm}. In addition, we consider \textbf{WavLM-large} to examine the effect of model capacity on downstream performance; in particular, we test whether the increased parameterization and representational power of a larger encoder translates into more informative embeddings and improved generalization.

\subsection{Pooling methods}
To aggregate the frame-level representations produced by the self-supervised models into a fixed-dimensional utterance-level representation, we compute temporal first- and second-order statistics of the embeddings, namely the mean and standard deviation across frames. Let $\mathbf{h}_t \in \mathbb{R}^d$ denote the frame-level embedding at time step $t$, for $t = 1, \dots, T$, where $d$ is the embedding dimension and $T$ is the number of frames. The temporally pooled mean representation is defined as
\[
\boldsymbol{\mu} = \frac{1}{T} \sum_{t=1}^{T} \mathbf{h}_t .
\]

The corresponding standard deviation vector is given by
\[
\boldsymbol{\sigma} = \sqrt{\frac{1}{T} \sum_{t=1}^{T} \left(\mathbf{h}_t - \boldsymbol{\mu}\right) \odot \left(\mathbf{h}_t - \boldsymbol{\mu}\right)},
\]
where $\odot$ denotes element-wise multiplication. When only mean pooling is applied, the resulting utterance-level representation remains in $\mathbb{R}^d$. When both mean and standard deviation are concatenated, the resulting representation lies in $\mathbb{R}^{2d}$
\[
\mathbf{z}_{\text{mean+std}} = [\boldsymbol{\mu}; \boldsymbol{\sigma}].
\]

To further investigate whether richer temporal structure can yield more robust representations for the task, we also employ covariance pooling (similar to \cite{kakouros2023speech}), which captures second-order dependencies across feature dimensions over time. Unlike mean and standard deviation pooling, which summarize each feature dimension independently, covariance pooling models the pairwise relationships among feature dimensions across the utterance. Given the centered frame-level embeddings $\mathbf{h}_t - \boldsymbol{\mu}$, the covariance matrix is defined as
\[
\mathbf{C} = \frac{1}{T} \sum_{t=1}^{T} \left(\mathbf{h}_t - \boldsymbol{\mu}\right)\left(\mathbf{h}_t - \boldsymbol{\mu}\right)^\top \in \mathbb{R}^{d \times d}.
\]

The resulting covariance matrix captures the temporal co-variation between embedding dimensions and therefore preserves information about the internal structure of the representation sequence that is not retained by marginal first- and second-order statistics alone.

\subsection{Classification}
For classification, we use a deep neural network (DNN) classifier consisting of a lightweight feed-forward network built on top of frozen \texttt{openSMILE} or SSL features. In this setup, the upstream encoder (e.g., \texttt{HuBERT} or \texttt{WavLM}) is used strictly as a fixed feature extractor and is never fine-tuned during downstream training. The classifier directly consumes a pooled feature representation of shape $[B, D]$, where $B$ denotes the batch size and $D$ the feature dimensionality.

The classifier head first projects the input representation to \texttt{projection\_dim} $= 512$ through a linear layer, followed by a nonlinear activation function (\texttt{GELU}) and dropout of $0.2$. This is followed by two additional hidden layers of \texttt{hidden\_dims} $= [256, 128]$, each again followed by activation and dropout, before a final linear layer produces logits for the two target classes, \texttt{ADS} and \texttt{CDS}. Model training is performed using cross-entropy loss and the \texttt{AdamW} optimizer, with default hyperparameters of learning rate $10^{-3}$ and weight decay $10^{-4}$. Optimization uses mini-batches with a batch size of $16$, together with gradient accumulation over $4$ steps, for $20$ epochs. After each epoch, performance on the development split is evaluated using training loss, development loss, accuracy, precision, recall, and F1 score. The checkpoint achieving the best development F1 score is retained, and early stopping is applied when development F1 fails to improve for $5$ epochs. After training, the best saved checkpoint is reloaded and used for final evaluation on the test split.

\subsection{Linear probes for model interpretability}
We train a set of \textit{linear probes} to assess the extent to which different representations encode ADS--CDS style-related information. Specifically, we probe both (i) the pooled SSL embeddings and (ii) the logits from the output layer of the ADS/CDS classifier network. For the latter, within each cross-validation fold, we train the ADS/CDS MLP classifier, save the classifier decision logit for every sample, and then run a separate per-feature Ridge regression using these logits as probe inputs. For each representation, we fit a separate linear regression model to predict the sentence-level (eGeMAPS) acoustic-prosodic features extracted with \texttt{openSMILE}, including, for example, the mean fundamental frequency (mean F0). This setup provides a controlled and interpretable evaluation of the information available in the representation. In particular, given an utterance-level representation $\mathbf{z} \in \mathbb{R}^d$, a linear probe predicts a target feature $y$ as
\[
\hat{y} = \mathbf{w}^\top \mathbf{z} + b,
\]
where $\mathbf{w} \in \mathbb{R}^d$ and $b \in \mathbb{R}$ are the probe parameters. In the case of predicting mean F0, $y$ corresponds to the reference mean F0 value extracted from the same utterance using \texttt{openSMILE}, and $\hat{y}$ is the corresponding estimate obtained from the probed representation. Probe performance is evaluated using only two regression metrics: Pearson correlation coefficient, which measures the linear association between the predicted and reference values, and the coefficient of determination ($R^2$), which measures the proportion of variance in the target feature explained by the probe predictions.

\section{Experiments}

\begin{figure}[t]
    \centering
    \includegraphics[width=\columnwidth]{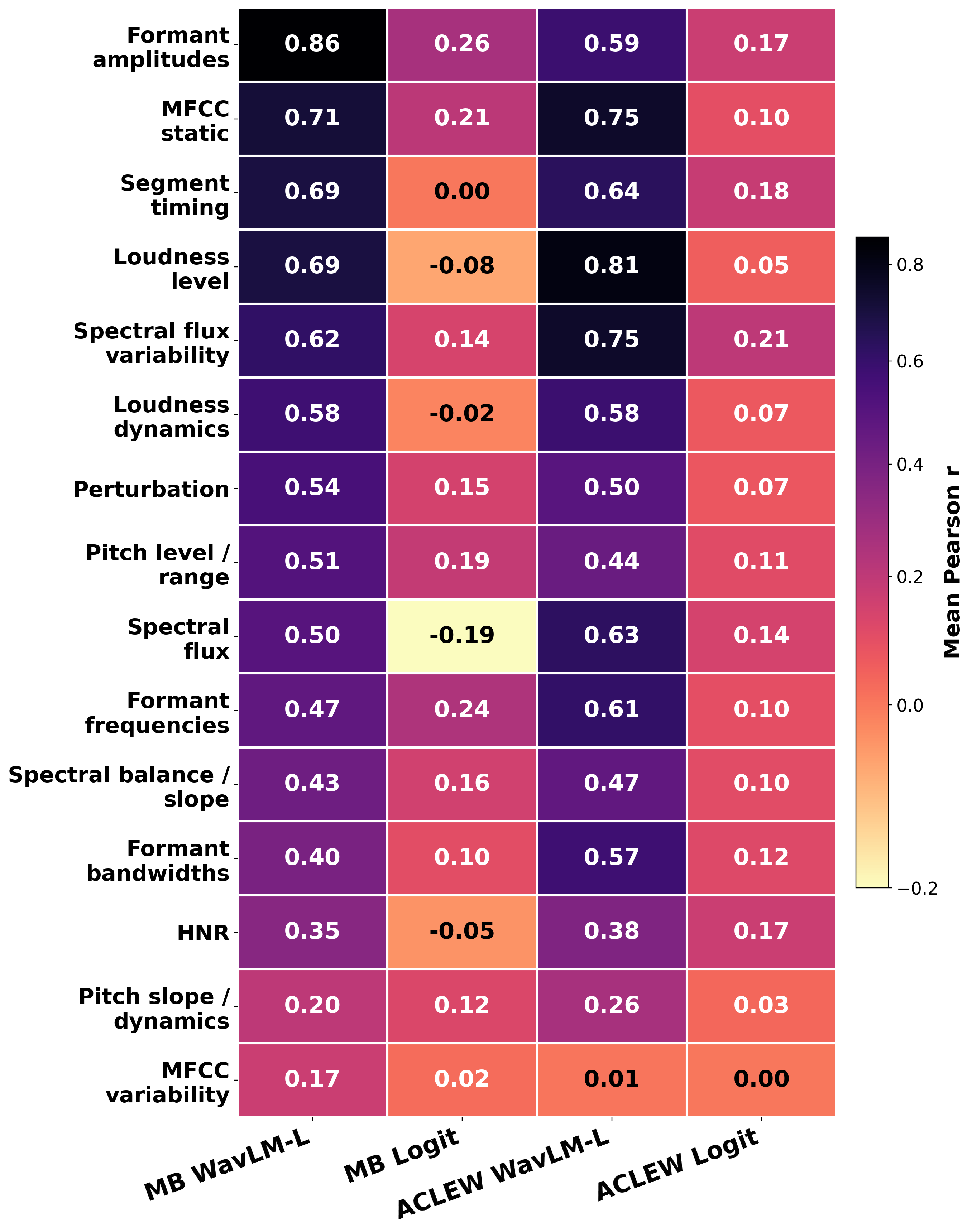}
    \caption{Mean Pearson \(r\) for grouped eGeMAPS features, computed from the correlations of WavLM-Large mean-pooled embeddings and logits with eGeMAPS for MB and ACLEW.}
    \label{fig:heatmap-single}
    \vspace{-6mm}
\end{figure}

We conducted experiments for all feature sets using a simple DNN classifier on both datasets (ManyBabies and ACLEW), evaluating all applicable pooling methods. For openSMILE, experiments were limited to the originally extracted acoustic descriptors, as no alternative pooling strategy was applicable. For WavLM-Base, WavLM-Base+, WavLM-Large, and HuBERT-Base, we evaluated mean, mean+std, and covariance pooling. Performance is reported in terms of unweighted accuracy and F1 score. All experiments were carried out using fixed data splits and fixed random seeds. Further details on the data partitioning are provided in the next subsection.

For the probing experiments, we trained ridge regression models using WavLM-Base+ and WavLM-Large representations obtained with mean and mean+std pooling. Covariance pooling was not considered in these experiments, as the number of unique covariance coefficients grows as $\frac{d(d-1)}{2}$, where $d$ denotes the embedding dimension---for example, for WavLM-Base+ with embedding dimension $d=768$, mean pooling yields a 768-dimensional vector, mean+std pooling yields a 1,536-dimensional vector, and covariance pooling yields $\frac{d(d-1)}{2}=294{,}528$ unique covariance coefficients. This leads to a rapid increase in feature dimensionality and, consequently, to a substantially more complex model (covariance pooling yields a representation approximately $383.5\times$ larger than mean pooling and $191.75\times$ larger than mean+std pooling). For example, when projecting the pooled representation to 512 dimensions, the first linear layer requires approximately $768\times512=0.39$ million parameters for mean pooling, $1{,}536\times512=0.79$ million for mean+std pooling, and $294{,}528\times512=150.8$ million for covariance pooling, excluding biases. Given the limited number of datapoints available, such a setting would likely result in unstable estimates.

\subsection{Data splits}

\begin{table}[t]
\centering
\caption{Five-fold cross-validation results for all runs of ManyBabies corpus. Values are reported as mean $\pm$ standard deviation. Best mean value in each metric is shown in bold.}
\label{tab:mb_classification_results}
\footnotesize
\setlength{\tabcolsep}{3pt}
\begin{tabular}{llcc}
\toprule
\textbf{Model} & \textbf{Pooling} & \textbf{Accuracy} & \textbf{F1} \\
\midrule
openSMILE-eGeMAPS & Mean        & 0.714 $\pm$ 0.073 & 0.788 $\pm$ 0.066 \\
\midrule
HuBERT-base       & Covariance        & 0.882 $\pm$ 0.056 & 0.909 $\pm$ 0.046 \\
HuBERT-base       & Mean        & 0.887 $\pm$ 0.027 & 0.914 $\pm$ 0.026 \\
HuBERT-base       & Mean+Std    & 0.889 $\pm$ 0.042 & 0.913 $\pm$ 0.040 \\
\midrule
WavLM-base+       & Covariance        & 0.901 $\pm$ 0.041 & 0.924 $\pm$ 0.035 \\
WavLM-base+       & Mean        & 0.890 $\pm$ 0.035 & 0.912 $\pm$ 0.033 \\
WavLM-base+       & Mean+Std    & 0.906 $\pm$ 0.054 & 0.926 $\pm$ 0.045 \\
\midrule
WavLM-base        & Covariance        & 0.867 $\pm$ 0.029 & 0.897 $\pm$ 0.026 \\
WavLM-base        & Mean        & 0.866 $\pm$ 0.036 & 0.896 $\pm$ 0.030 \\
WavLM-base        & Mean+Std    & 0.865 $\pm$ 0.041 & 0.894 $\pm$ 0.040 \\
\midrule
WavLM-large       & Covariance        & 0.892 $\pm$ 0.061 & 0.919 $\pm$ 0.048 \\
WavLM-large       & Mean        & 0.908 $\pm$ 0.040 & 0.928 $\pm$ 0.034 \\
WavLM-large       & Mean+Std    & \textbf{0.917 $\pm$ 0.046} & \textbf{0.933 $\pm$ 0.040} \\
\bottomrule
\end{tabular}
\vspace{-2mm}
\end{table}

For the ACLEW corpus, we used only the English-language subset, which comprises four distinct datasets collected at different recording locations. This structure provides a natural and meaningful basis for data partitioning; accordingly, recording location was used as the splitting criterion. This design additionally enables evaluation in a cross-corpus generalization setting. In total, four splits were defined. In each split, the training and development sets comprised data from three recording locations, while the remaining location was held out as the test set.

For the ManyBabies corpus, we constructed splits based on recording ID while maintaining a balanced overall distribution of ADS and CDS across subsets. More specifically, we defined five fixed splits in which 80\% of the data, partitioned by BabyID (i.e., caregiver--infant family), was assigned to the training and development sets, and the remaining 20\% was reserved for testing. This ensured that no BabyID appeared in both the training/development and test sets.

\section{Results and Discussion}

\begin{table}[t]
\centering
\caption{Four-fold cross-validation results for all runs for ACLEW dataset. Values are reported as mean $\pm$ standard deviation. Best mean value in each metric is shown in bold.}
\label{tab:aclew_classification_results}
\footnotesize
\setlength{\tabcolsep}{3pt}
\begin{tabular}{llcc}
\toprule
\textbf{Model} & \textbf{Pooling} & \textbf{Accuracy} & \textbf{F1} \\
\midrule
openSMILE-eGeMAPS & Mean        & 0.551 $\pm$ 0.076 & 0.154 $\pm$ 0.267 \\
\midrule
HuBERT-base       & Covariance        & 0.602 $\pm$ 0.018 & 0.591 $\pm$ 0.062 \\
HuBERT-base       & Mean        & 0.606 $\pm$ 0.021 & 0.600 $\pm$ 0.039 \\
HuBERT-base       & Mean+Std    & 0.594 $\pm$ 0.021 & 0.606 $\pm$ 0.046 \\
\midrule
WavLM-base+       & Covariance        & 0.616 $\pm$ 0.020 & 0.603 $\pm$ 0.031 \\
WavLM-base+       & Mean        & 0.621 $\pm$ 0.009 & 0.615 $\pm$ 0.036 \\
WavLM-base+       & Mean+Std    & 0.611 $\pm$ 0.026 & 0.628 $\pm$ 0.050 \\
\midrule
WavLM-base        & Covariance        & 0.600 $\pm$ 0.032 & 0.597 $\pm$ 0.039 \\
WavLM-base        & Mean        & 0.613 $\pm$ 0.019 & 0.588 $\pm$ 0.047 \\
WavLM-base        & Mean+Std    & 0.606 $\pm$ 0.009 & 0.595 $\pm$ 0.052 \\
\midrule
WavLM-large       & Covariance        & 0.641 $\pm$ 0.031 & 0.610 $\pm$ 0.032 \\
WavLM-large       & Mean        & \textbf{0.644 $\pm$ 0.015} & 0.658 $\pm$ 0.051 \\
WavLM-large       & Mean+Std    & 0.642 $\pm$ 0.034 & \textbf{0.660 $\pm$ 0.050} \\
\bottomrule
\end{tabular}
\vspace{-6mm}
\end{table}

\subsection{Classification and pooling}

\begin{table*}[t]
\centering
\footnotesize
\setlength{\tabcolsep}{4.5pt}
\renewcommand{\arraystretch}{1.08}
\caption{Top-10 openSMILE features for ManyBabies corpus by logit Pearson \(r\), with embedding \(r\) in parentheses. Highlighted values are the top-5 embeddings \(r\) per column.}
\label{tab:mb_top10_logits_vs_embeddings_both_models}
\begin{tabular}{lcccc}
\toprule
& \multicolumn{2}{c}{\textbf{WavLM-Base-Plus}} & \multicolumn{2}{c}{\textbf{WavLM-Large}} \\
\cmidrule(lr){2-3}\cmidrule(lr){4-5}
\textbf{Feature} & \textbf{m} & \textbf{m+s} & \textbf{m} & \textbf{m+s} \\
\midrule
\texttt{mfcc2\_mean}  &
0.505 (\cellcolor{black!12}0.791) &
0.523 (\cellcolor{black!12}0.771) &
0.522 (\cellcolor{black!12}0.799) &
0.536 (\cellcolor{black!12}0.771) \\

\texttt{f3freq\_mean} &
0.429 (0.543) &
0.437 (0.533) &
0.431 (0.562) &
0.451 (0.512) \\

\texttt{f3amp\_std}  &
0.331 (\cellcolor{black!15}\textbf{0.806}) &
0.382 (\cellcolor{black!5}0.671) &
0.354 (\cellcolor{black!15}\textbf{0.840}) &
0.362 (\cellcolor{black!15}\textbf{0.778}) \\

\texttt{arUV\_mean}   &
0.358 (0.559) &
0.389 (\cellcolor{black!3}0.636) &
0.345 (0.648) &
0.365 (\cellcolor{black!3}0.634) \\

\texttt{f0p80}     &
0.390 (\cellcolor{black!3}0.666) &
0.402 (\cellcolor{black!15}\textbf{0.804}) &
0.402 (\cellcolor{black!5}0.732) &
0.413 (\cellcolor{black!7}0.745) \\

\texttt{mfcc2V\_mean} &
0.376 (\cellcolor{black!10}0.700) &
0.393 (0.562) &
0.399 (\cellcolor{black!3}0.692) &
0.404 (\cellcolor{black!5}0.697) \\

\texttt{hammUV\_mean} &
0.323 (0.578) &
0.357 (\cellcolor{black!7}0.680) &
0.300 (0.612) &
0.325 (\cellcolor{black!10}0.643) \\

\texttt{arV\_mean}    &
0.325 (\cellcolor{black!7}0.687) &
0.358 (0.570) &
0.341 (\cellcolor{black!7}0.744) &
0.350 (\cellcolor{black!1}0.670) \\

\texttt{h1h2\_mean}   &
0.294 (0.503) &
0.307 (0.482) &
0.346 (0.582) &
0.344 (0.534) \\

\texttt{f2freq\_mean} &
0.289 (0.576) &
0.300 (0.600) &
0.319 (0.622) &
0.326 (0.540) \\
\bottomrule
\end{tabular}
\end{table*}

\begin{table*}[t]
\centering
\footnotesize
\setlength{\tabcolsep}{4.5pt}
\renewcommand{\arraystretch}{1.08}
\caption{Top-10 openSMILE features for ACLEW corpus by logit Pearson \(r\), with embedding \(r\) in parentheses. Highlighted values are the top-5 embedding \(r\) per column.}
\label{tab:aclew_top10_logits_vs_embeddings_both_models}
\resizebox{\columnwidth}{!}{%
\begin{tabular}{lcccc}
\toprule
& \multicolumn{2}{c}{\textbf{WavLM-Base-Plus}} & \multicolumn{2}{c}{\textbf{WavLM-Large}} \\
\cmidrule(lr){2-3}\cmidrule(lr){4-5}
\textbf{Feature} & \textbf{m} & \textbf{m+s} & \textbf{m} & \textbf{m+s} \\
\midrule
\texttt{hnr\_mean} &
0.405 (\cellcolor{black!12}\textbf{0.672}) &
0.461 (\cellcolor{black!12}\textbf{0.650}) &
0.342 (\cellcolor{black!12}\textbf{0.757}) &
0.461 (\cellcolor{black!12}\textbf{0.742}) \\

\texttt{f3amp\_mean} &
0.324 (\cellcolor{black!12}\textbf{0.775}) &
0.388 (\cellcolor{black!12}\textbf{0.762}) &
0.244 (\cellcolor{black!12}\textbf{0.843}) &
0.366 (\cellcolor{black!12}\textbf{0.832}) \\

\texttt{f1amp\_mean} &
0.317 (\cellcolor{black!12}\textbf{0.774}) &
0.382 (\cellcolor{black!12}\textbf{0.761}) &
0.238 (\cellcolor{black!12}\textbf{0.835}) &
0.358 (\cellcolor{black!12}\textbf{0.823}) \\

\texttt{f2amp\_mean} &
0.317 (\cellcolor{black!12}\textbf{0.776}) &
0.382 (\cellcolor{black!12}\textbf{0.763}) &
0.238 (\cellcolor{black!12}\textbf{0.843}) &
0.360 (\cellcolor{black!12}\textbf{0.832}) \\

\texttt{shimdb\_stdn} &
0.306 (\cellcolor{black!12}\textbf{0.669}) &
0.385 (\cellcolor{black!12}\textbf{0.668}) &
0.218 (0.707) &
0.351 (0.709) \\

\texttt{fluxv\_stdn} &
0.270 (0.647) &
0.341 (0.640) &
0.210 (\cellcolor{black!12}\textbf{0.723}) &
0.331 (\cellcolor{black!12}\textbf{0.713}) \\

\texttt{f0st\_p50} &
0.260 (0.589) &
0.310 (0.574) &
0.211 (0.701) &
0.309 (0.682) \\

\texttt{f0st\_mean} &
0.259 (0.593) &
0.311 (0.577) &
0.211 (0.705) &
0.310 (0.687) \\

\texttt{f0st\_p20} &
0.260 (0.567) &
0.306 (0.549) &
0.216 (0.679) &
0.308 (0.661) \\

\texttt{fluxuv\_mean} &
0.234 (0.575) &
0.306 (0.579) &
0.229 (0.696) &
0.317 (0.679) \\
\bottomrule
\end{tabular}%
}
\end{table*}

The results of the pooling experiments are presented in Tables~\ref{tab:mb_classification_results} and~\ref{tab:aclew_classification_results}. We observe that all SSL representations clearly outperform the 88-dimensional eGeMAPS baseline. This difference is not unexpected, given the substantially greater representational capacity of SSL models: whereas eGeMAPS provides a compact, hand-crafted representation of acoustic and prosodic characteristics, SSL representations are high-dimensional features learned directly from large-scale speech data and can encode richer information spanning acoustic, phonetic, and linguistic content. We also find a substantial performance gap between the two datasets, with ManyBabies achieving much higher scores (best: 91.7\% accuracy and 93.3\% F1) than ACLEW (64.4\% accuracy and 66.0\% F1). This difference is likely due to differences in recording quality: ManyBabies consists of high-quality recordings, whereas ACLEW contains naturalistic day-long recordings, which are inherently much noisier.

We also observe some differences across model variants, although the differences between model families (HuBERT vs.\ WavLM) are less pronounced, at least for the base models. The use of WavLM Base+ with its denoising-based objective leads to improved performance, even though the model capacity remains unchanged. In contrast, the large model outperforms all other variants and achieves the best overall performance.

With respect to the pooling strategies, no substantial differences are observed, as all three approaches (mean, mean+std, and covariance) yield broadly comparable results. However, covariance pooling is considerably less efficient, since the number of unique pairwise covariance coefficients grows quadratically with the embedding dimension, as $\frac{d(d-1)}{2}=\mathcal{O}(d^2)$, compared with the linear dimensionality of mean and mean+std pooling. In our experiments, this additional complexity did not lead to a corresponding improvement in classification performance, as covariance pooling produced results very similar to those obtained with the simpler pooling strategies. This suggests that the pairwise relationships captured by covariance pooling contributed little additional discriminative information for the ADS--CDS task beyond that already represented by the mean and mean+std.

\subsection{Linear Probing}

The linear probing experiments provide insight into (i) the acoustic information encoded in the SSL embeddings and (ii) the subset of that information retained by the classifier for the task when operating over the same embeddings. An overview of the results is presented in Tables \ref{tab:mb_top10_logits_vs_embeddings_both_models} and \ref{tab:aclew_top10_logits_vs_embeddings_both_models} (abbrev: \textit{arUV\_m}=mean unvoiced alpha ratio), as well as in Figure \ref{fig:heatmap-single}.

Across both corpora, the clearest pattern is that correlations with the embeddings are consistently much stronger than correlations with the logits. This suggests that the SSL representations preserve substantially richer acoustic information than what is reflected in the final prediction layer. This reduction is expected because the final logits constitute a highly compressed, task-specific representation. Nevertheless, the comparison is important because it allows us to identify which eGeMAPS features remain strongly represented in the SSL embeddings, which associations are substantially attenuated at the decision level, and whether these feature-specific changes are consistent across ManyBabies and ACLEW.

In particular, in ManyBabies, logit \(r\) values are moderate (approx. 0.29--0.54), with the strongest associations concentrated in MFCC, formant, and pitch-related features---most notably \texttt{mfcc2\_mean}, which is the highest-ranked logit feature in all four settings. By contrast, embedding correlations frequently reach the 0.67--0.84 range, with particularly large increases for \texttt{f3amp\_std}, \texttt{mfcc2\_mean}, and \texttt{f0p80}. In ACLEW, this gap is even more pronounced: logit correlations are generally lower (about 0.21--0.46), whereas embedding correlations remain high (approx. 0.55--0.84). This indicates that the models encode robust low-level acoustic structure even when that information is only weakly expressed in the final logits. 

The strongest ACLEW predictors are dominated by voice quality and spectral amplitude measures, including \texttt{hnr\_mean}, \texttt{f1/2/3amp\_mean}, \texttt{shimdb\_stdn}, and \texttt{fluxv\_stdn}. These results indicate clear corpus-specific differences in acoustic salience. One likely contributing factor is the difference in recording conditions across corpora. ManyBabies recordings are comparatively more controlled and studio-like, whereas ACLEW contains more naturalistic, daylong recordings with greater variation in background noise, talker distance, reverberation, overlap, and different recording device characteristics (compared to ManyBabies). This type of environmental differences can affect not only broad spectral or noise-related measures, but also specific acoustic feature families, including formant amplitudes and spectral flux. For example, variation in microphone placement or vocal effort may change the relative amplitudes of formant regions, while background noise and reverberation can influence measures of spectral stability. Thus, the stronger presence of amplitude- and voice-quality-related features in ACLEW may reflect both differences in the speech being captured and differences introduced by the naturalistic recording environment itself.

To provide a broader overview, the 88 OpenSMILE features were grouped into a smaller set of acoustically meaningful categories using a rule-based mapping derived from the feature names (Figure \ref{fig:heatmap-single}). The same general pattern emerges in this grouped analysis: the SSL embeddings clearly encode substantial acoustic information, but the extent to which that information is retained differs across the two datasets.

\section{Conclusions}

In this work, we presented a comprehensive analysis of classification performance for distinguishing CDS from ADS using two complementary datasets: ManyBabies, which consists of more controlled in-lab recordings, and ACLEW, which consists of naturalistic long-form audio. In addition to evaluating classification performance, we examined the interpretability of self-supervised learning (SSL) representations with respect to OpenSMILE acoustic features, focusing on both the acoustic information encoded in the embeddings and the extent to which this information is taken up by the classifier in the CDS–ADS distinction task.

Overall, the results show that SSL representations provide a clear advantage over the eGeMAPS baseline, with the strongest performance achieved by larger models and models enhanced through denoising, while differences between pooling strategies remain relatively small. At the same time, the linear probing analysis reveals that SSL embeddings retain substantially richer acoustic information than is ultimately reflected in the classifier’s final decision layer. This suggests that although SSL models encode a broad range of low-level and higher-level acoustic cues, only a subset of these cues is emphasized by the task-specific classifier. The results also point to consistent corpus-specific differences in the acoustic cues that are most salient, likely reflecting both differences in speech style and differences in recording conditions between controlled in-lab and naturalistic long-form audio.

\section*{Acknowledgment}

SK and DK were supported by the L-SCALE project funded by Kone Foundation. DK was also funded by Research Council of Finland (grant no. 369379). SK was also funded by Research Council of Finland (grant no. 365282). The authors wish to acknowledge CSC – IT Center for Science, Finland, for providing the computational resources.

\subsection*{AI-Generated Content Disclosure}
OpenAI Codex with GPT-5.5 was used to assist with table formatting. AI tools were not used to design or run the experiments, compute the reported metrics, or interpret experimental results.

\bibliographystyle{IEEEtran}
\bibliography{mybib}

\end{document}